\documentclass[prl,aps,twocolumn,showpacs,citeautoscript,reprint,superscriptaddress]{revtex4-1}
\usepackage{graphicx}
\usepackage{bm,amsmath,amssymb,wasysym,amsfonts,dcolumn}
\usepackage[colorlinks=true, urlcolor=blue, linkcolor=blue, citecolor=blue, pdfborder={0 0 0}]{hyperref}
\usepackage[usenames,dvipsnames]{xcolor}
\usepackage{mathrsfs}
\usepackage{booktabs}
\usepackage{multirow}
\usepackage{makecell}

\usepackage{hyperref}

\newcommand{\DOI}[1]{\href{https://doi.org/#1}}

\begin{document}

\title{Orbital-excitation-dominated magnetization dissipation and quantum oscillation of Gilbert damping in Fe films}

\author{Yue Chen}
\altaffiliation{These authors contributed equally to this study.}
\affiliation{Center for Advanced Quantum Studies and Department of Physics, Beijing Normal University, Beijing 100875, China}
\affiliation{Institute for Nanoelectronics and Quantum Computing, Fudan University, Shanghai 200433, China}
\affiliation{Interdisciplinary Center for Theoretical Physics and Information Sciences (ICTPIS), Fudan University, Shanghai 200433, China}
\author{Haoran Chen}
\altaffiliation{These authors contributed equally to this study.}
\author{Xi Shen}
\affiliation{Department of Physics and State Key Laboratory of Surface Physics, Fudan University, Shanghai 200433, China}
\author{Weizhao Chen}
\affiliation{Center for Advanced Quantum Studies and Department of Physics, Beijing Normal University, Beijing 100875, China}
\affiliation{Institute for Nanoelectronics and Quantum Computing, Fudan University, Shanghai 200433, China}
\affiliation{Interdisciplinary Center for Theoretical Physics and Information Sciences (ICTPIS), Fudan University, Shanghai 200433, China}
\author{Yi Liu}
\affiliation{Institute for Quantum Science and Technology, Department of Physics, Shanghai University, Shanghai 200444, China}
\author{Yizheng Wu}
\email{wuyizheng@fudan.edu.cn}
\affiliation{Department of Physics and State Key Laboratory of Surface Physics, Fudan University, Shanghai 200433, China}
\affiliation{Shanghai Research Center for Quantum Sciences, Shanghai 201315, China}
\affiliation{Shanghai Key Laboratory of Metasurfaces for Light Manipulation, Fudan University, Shanghai 200433, China}
\author{Zhe Yuan}
\email{yuanz@fudan.edu.cn}
\affiliation{Institute for Nanoelectronics and Quantum Computing, Fudan University, Shanghai 200433, China}
\affiliation{Interdisciplinary Center for Theoretical Physics and Information Sciences (ICTPIS), Fudan University, Shanghai 200433, China}
\date{\today}

\begin{abstract}
Using first-principles electronic structure calculation, we demonstrate the spin dissipation process in bulk Fe by orbital excitations within the energy bands of pure spin character. The variation of orbitals in the intraband transitions provides an efficient channel to convert spin to orbital angular momentum with spin-orbit interaction. This mechanism dominates the Gilbert damping of Fe below room temperature. The theoretical prediction is confirmed by the ferromagnetic resonance experiment performed on single-crystal Fe(001) films. A significant thickness-dependent damping oscillation is found at low temperature induced by the quantum well states of the corresponding energy bands. Our findings not only explain the microscopic nature of the recently reported ultralow damping of Fe-based alloys, but also help for the understanding of the transport and dissipation process of orbital currents.
\end{abstract}
\maketitle

Gilbert damping is a phenomenological parameter to describe the dissipation process of magnetization dynamics using the Landau-Lifshitz-Gilbert equation~\cite{Urban:prl01,Malinowski:apl09,Bonetti:prl16,Li:prl19,Khodadadi:prl20}. Analogous to the viscous damping in a harmonic oscillator, the magnitude of Gilbert damping determines the magnetic switching time~\cite{Sun:prb00}, the critical current density required for switching using spin torques~\cite{Brataas:nm12}, and the velocity of current-driven magnetic textures~\cite{Tatara:pr08,Wang:ne20}. Thus, understanding and controlling Gilbert damping is highly desired to improve the performance of spintronics devices with low energy consumption. 

\begin{figure}[b]
\centering
\includegraphics[width=\columnwidth]{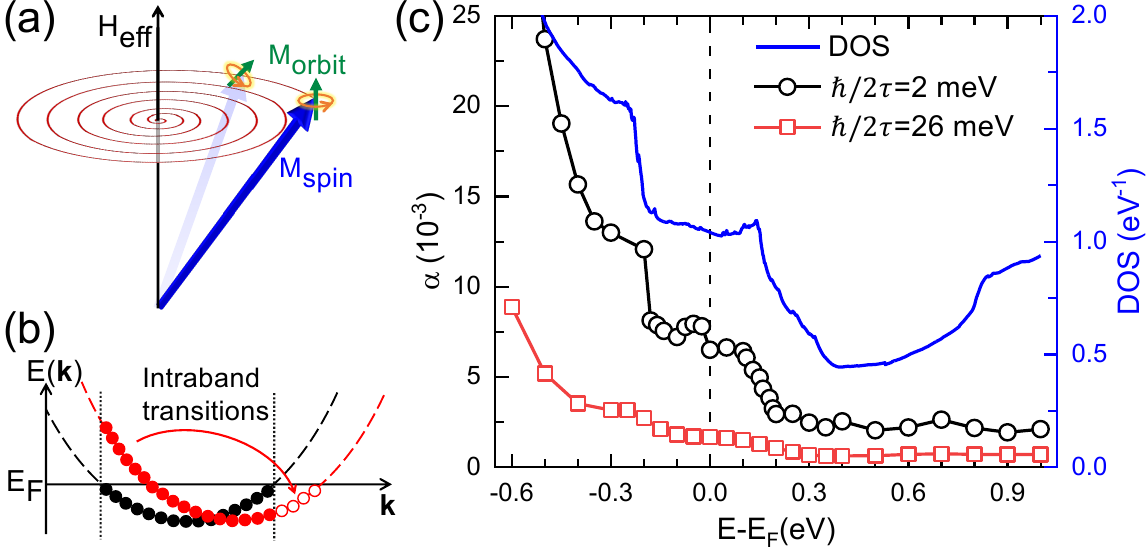}
\caption{(a) Schematic illustration of magnetic damping. The orbital moments (green arrows) are excited and varying during magnetization precession. Then spin (blue arrows) is converted to orbital angular momentum via SOC leading to the magnetic damping. (b) Sketch of the intraband transition. Equilibrium occupation of an energy band (black) evolves out of equilibrium (red) with magnetization precession until intraband transitions bring it back to equilibrium again. (c) Calculated damping of bcc Fe using Eq.~\eqref{eq:tcm} with low (black circles) and high (red squares) scattering rate of conduction electrons. The calculated DOS is plotted (blue line) for comparison.}
\label{fig1}
\end{figure}

There is a general consensus that Gilbert damping in metallic ferromagnets arises from the scattering of electrons at the Fermi level during magnetization precession~\cite{Kambsersky:CJP76,Tserkovnyak:rmp05,Scheck:prl07,Brataas:PRL08,Ebert:PRL11,Liu:prl14,Schoen:prb17,Starikov:prb18}, as sketched in Fig.~\ref{fig1}(a). The presence of spin-orbit coupling (SOC) allows spin flipping in the scattering resulting in the dissipation of spin angular momentum and hence the magnetization dissipation occurs because electronic spin is the main source of magnetization in solids. In strongly disordered samples, the damping usually increases with increasing the scattering rate due to the interband transitions. In high-quality single crystals, the damping is found to increase with decreasing the electron scattering rate, which is well described by the breathing Fermi surface (BFS) model with the intraband transition~\cite{Kambersky:cjp70}; see Fig.~\ref{fig1}(b). Nevertheless, the BFS model only explains the energy loss without a detailed picture of angular momentum dissipation. It is well known that orbital angular momentum also contributes to magnetization, and the recently discovered effect of orbital current on the spin-dependent transport has attracted great attention in the field of spintronics~\cite{Ding:prl22,Choi:nat23,Seifert:natnano23}. However, there is still lack of discussion about the effect of the orbital angular momentum on the magnetization dissipation process. A comprehensive understanding of the orbital effect in the Gilbert damping would be also helpful for gaining a deep insight into the transport and dynamics of orbital currents~\cite{Lyalin:prl23,Sala:prl23,Sohn:prl24,Rang:prb24}.

In this Letter, we investigate the contribution of each energy band for Gilbert damping of bcc Fe via first-principles calculations, and reveal the dominant role of orbital excitations. Intraband transitions in the energy bands with pure spin and hybridized orbitals alter orbital angular momentum, converting spin angular momentum to the orbital degree of freedom via SOC. The orbital angular momentum then dissipates to lattice through electron-phonon interaction. This is essentially the dominant mechanism contributing to Gilbert damping in Fe, attributed to the small hole pockets of $\Delta_5$ character near the H points in the Brillouin zone (BZ). To validate our theoretical predictions, we perform ferromagnetic resonance (FMR) experiment on ultrathin Fe(001) films with continuously varying thicknesses. The measured damping at low temperature exhibits quantum oscillations with the period of approximately 9 atomic layers, in excellent agreement with the quantum well states (QWSs) of $\Delta_5^\uparrow$ bands. Our findings reveal the microscopic nature of the recently reported ultralow damping in Fe-based alloys. The understanding of orbital excitations in the Gilbert damping process sheds light on the transport and dynamical properties of orbital angular momentum.

%%%%%%%%10%%%%%%%%20%%%%%%%%30%%%%%%%%40%%%%%%%%50%%%%%%%%60%%%%%%%%70%%%%%%%%80
{\it\color{red}Energy-dependent damping of Fe.---}Based upon first-principles electronic structure, the Gilbert damping can be calculated using Kambersk{\'y}'s torque-correlation model (TCM)~\cite{Kambsersky:CJP76,Gilmore:PRL07,Gilmore:JAP08} as
\begin{equation}
\alpha=\frac{\pi\hbar\gamma}{\mu_{0}M_s}\sum_{n,m}\int\frac{d^3k}{8\pi^3}\vert\Gamma_{nm}^{-}(\mathbf k)\vert^2\int dEA_{n\mathbf{k}}(E)A_{m\mathbf{k}}(E)\eta(E).\label{eq:tcm}
\end{equation}
Here the spectral function $A_{n\mathbf{k}}(E)$ is a Lorentzian centered at the band energies $E_{n\mathbf k}$ with its width corresponding to the scattering rate $\hbar/2\tau$~\cite{Gilmore:JAP08}. $\eta(E)=-\partial f/\partial E$ is the energy derivative of the Fermi-Dirac distribution, restricting the main contribution near the Fermi level $E_F$. The matrix element $\Gamma_{nm}^{-}(\mathbf k)=\langle\psi_{n\mathbf k}\vert[\sigma_-,H_{\rm so}]\vert\psi_{m\mathbf k}\rangle$ describes a torque on electron spin. $H_{\rm so}$ is the SOC Hamiltonian and $\sigma_-=\sigma_x-i\sigma_y$ is defined under the local quantization axis. 

By shifting $E_F$ in Eq.~\eqref{eq:tcm}, the calculated Gilbert damping $\alpha$ of Fe is plotted in Fig.~\ref{fig1}(c) with the scattering rate $\hbar/2\tau=2$~meV and 26~meV corresponding to the calculated $\rho=0.8~\mu\Omega\,{\rm cm}$ and $9.5~\mu\Omega\,{\rm cm}$ at $E_F$, respectively. The latter is comparable to the experimental resistivity of 9.61~$\mu\Omega$~cm at 293 K~\cite{Ho:JPS83}. Some previous studies in literature suggests that $\alpha$ is positively correlated to the total density of states (DOS) at $E_F$~\cite{Schoen:NP16,Chen:natphys18,Chen:prl23,Xu:prb19}, but uncorrelated damping and DOS has also been reported~\cite{Hou:prapplied19,Jiang:prb24,Bera:arxiv24}. Here the calculated DOS is shown in Fig.~\ref{fig1}(c) for comparison. In the energy range of $E-E_F<0.2$~eV, the calculated $\alpha$ with $\hbar/2\tau=2$~meV shows a similar trend as DOS. But this correlation does not hold at $E-E_F>0.5$ eV, where the DOS exhibits a remarkable increase but $\alpha$ is nearly unchanged. With $\hbar/2\tau=26$~meV, the calculated $\alpha$ is also nearly a constant at $E-E_F>0.5$~eV. Thus, the energy-dependent Gilbert damping can not be simply attributed to the DOS. In Fig.~\ref{fig1}(c), $\alpha$ with the low scattering rate is significantly larger than that with the high value indicating that the intraband contribution is dominant below room temperature. It is in agreement with the experimentally measured $\alpha$, which decreases monotonically from low to room temperature~\cite{Khodadadi:prl20}. Therefore, we focus on the intraband transition in the following and identify the band-resolved contribution.

{\it\color{red}Band-resolved damping.---}The intraband contribution can be directly calculated by choosing the same band ($m=n$) in Eq.~\eqref{eq:tcm}, where $\vert\Gamma_{nn}^{-}(\mathbf k)\vert^2$ is related to the dependence of band energy on magnetization orientation. Specifically, for $\mathbf{M}=M_s\hat x$, one has~\cite{Gilmore:JAP08},
\begin{equation}
\vert\Gamma_{nn}^{-}(\mathbf k)\vert^2={\left(\frac{\partial E_{n\mathbf k}}{\partial \theta_y} \right)}^2+{\left( \frac{\partial E_{n\mathbf k}}{\partial \theta_z} \right)}^2,\label{eq:Gamma}
\end{equation}	
where $\theta_{y}$ ($\theta_{z}$) is the small precessional angle of the magnetization deviating from the $\hat x$ axis towards $\hat y$ ($\hat z$). 

\begin{figure}[t]
\centering
\includegraphics[width=\columnwidth]{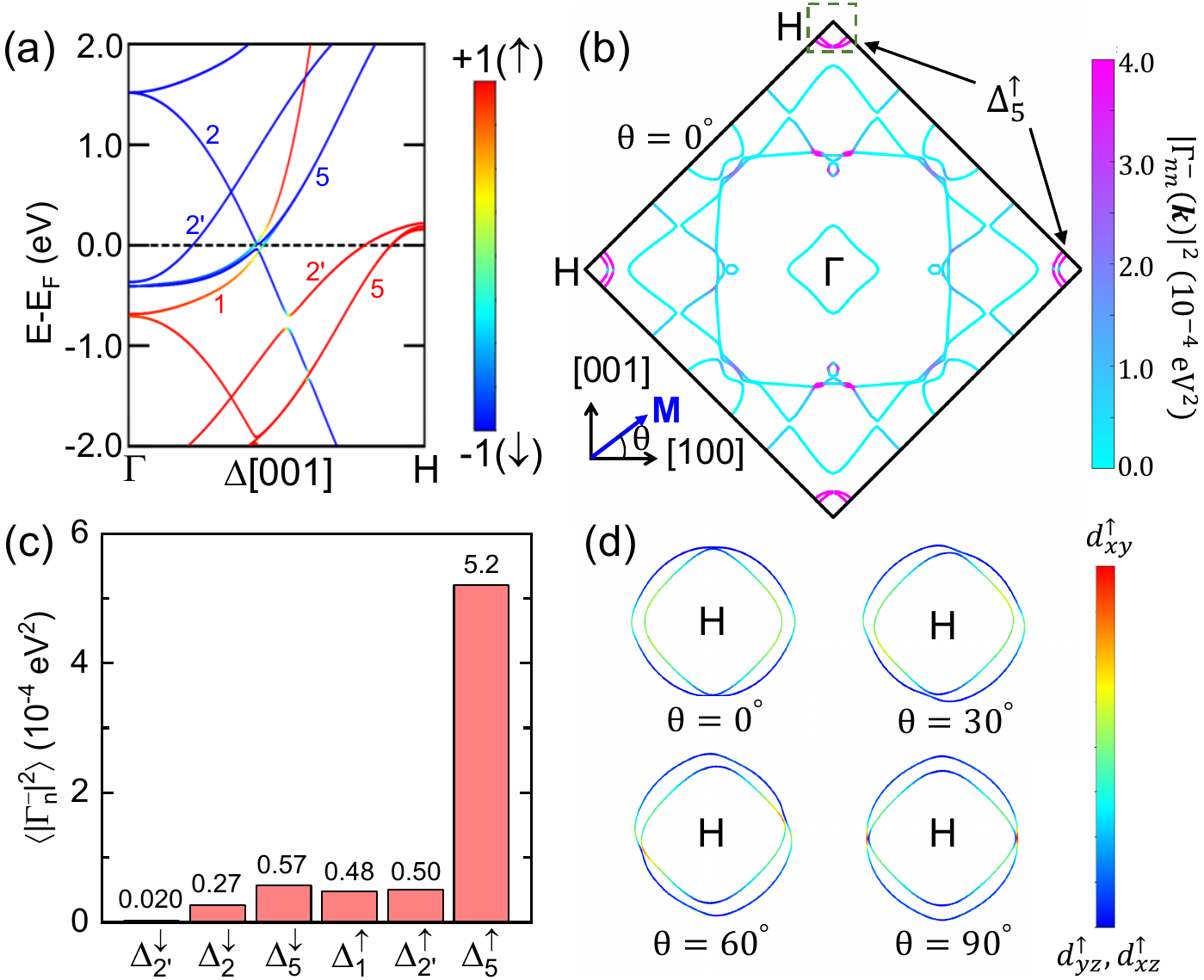}
\caption{(a) Calculated energy bands of Fe along $\Gamma$H. The color indicates the spin component projected onto the quantization axis and the number on every band identifies the corresponding $\Delta$-symmetry. (b) $k$-resolved intraband matrix element $\vert\Gamma^{-}_{nn}(\mathbf k)\vert^2$ in the (010) plane of the BZ with $\mathbf M\parallel$[100] ($\theta=0^\circ$). $\theta$ represents the angle between magnetization and the crystal axis [100]. (c) Averaged intraband matrix element over the Fermi surface for each band.  (d) Orbital component $d^\uparrow_{xy}$ ($l_z=\pm2$) and $d^\uparrow_{xz(yz)}$ ($l_z=\pm1$) in the $\Delta^\uparrow_5$ bands at $E_F$ near the H point in the dashed frame in panel (b).}
\label{fig2}
\end{figure}

The band structure of bulk Fe along [100] is plotted in Fig.~\ref{fig2}(a), where six categories of energy bands cross the Fermi level. To identify the contribution of each band, we explicitly compute $\vert\Gamma_{nn}^{-}(\mathbf k)\vert^2$ using Eq.~\eqref{eq:Gamma} and plot the $\mathbf k$-resolved results in (010) plane ($k_y=0$) of the BZ in Fig.~\ref{fig2}(b). The magnitude of $\vert\Gamma_{nn}^{-}(\mathbf k)\vert^2$ is not the same for every $k$ point and the maximum magnitude appears on the hole pockets around the H point. For each category, the averaged matrix element over the corresponding Fermi sheet $\langle\vert\Gamma^-_n\vert^2\rangle$ is plotted in Fig.~\ref{fig2}(c). Although the $\Delta_5^\uparrow$ bands only have small Fermi sheets near H points, they have the surprisingly dominant contribution, which is an order of magnitude larger than any other bands. The dominance of the $\Delta_5^\uparrow$ bands is independent of the specific orientation of magnetization~\cite{SM}.

Identifying the dominant contribution of Gilbert damping in Fe sheds light on the microscopic nature of the ultralow $\alpha$ of Fe$_{75}$Co$_{25}$~\cite{Schoen:NP16,Lee:NC17}, Fe$_{75}$Al$_{25}$~\cite{Wei:SAv21} and Fe$_{85}$V$_{15}$ alloys~\cite{Arora:prapplied21} observed in recent experiments. We calculated the spectral functions of these Fe-based alloys and discovered that doping Co, Al, or V results in the full occupation of the $\Delta_5^\uparrow$ bands~\cite{SM}. Thus these bands are shifted downwards away from $E_F$ and their contribution is significantly reduced, leading to the ultralow values of $\alpha$. 

{\it\color{red}Orbital-excitations-dominated Gilbert damping.---}The dominant role of the $\Delta_5^\uparrow$ bands can be understood as follows. The torque $\Gamma_{nn}^{-}(\mathbf k)\propto\langle\psi_{n\mathbf k}\vert l_z\sigma_--l_- \sigma_z\vert\psi_{n\mathbf k}\rangle$ requires the mixed components with either different spins or different magnetic quantum numbers $l_z$ in a single band. The $\Delta_5^\uparrow$ bands have pure majority-spin character at equilibrium, as shown in Fig.~\ref{fig2}(a). In contrast, the orbital characters of these states are strongly hybridized. The orbital components of $d_{xy}$ with $l_z=\pm2$ and $d_{xz(yz)}$ with $l_z=\pm1$ are plotted in Fig.~\ref{fig2}(d) with the magnetization rotating from [100] to [001]. It suggests that with magnetization precession, the orbital components keep varying and this quick variation allows reversible conversion between orbital and spin components of these nonequilibrium states through SOC~\cite{Ning:arxiv23}. In another word, the magnetization-precession-induced orbital excitations provide an efficient channel to convert spin to orbital angular momentum. It is worth noting that spin mixing occurs in the $\Delta_1^\uparrow$ band near $E_F$, but the calculated $\langle\vert\Gamma^-_n\vert^2\rangle$ is relatively small because $\Delta_1^\uparrow$ band mainly consists of $s$ and $d_{z^2}$ components with $l_z=0$ for both. Therefore, the Gilbert damping of Fe is dominated by the orbital excitations of $\Delta_5^\uparrow$ bands. At $E-E_F>0.5$~eV in Fig.~\ref{fig1}(c), the increase of DOS arises mainly from the $\Delta_2^\downarrow$ band~\cite{Singh:prb75,Callaway:prb77}, which does not show positive correlation with the calculated $\alpha$. The excited nonequilibrium orbital moments could potentially be detected via X-ray magnetic circular dichroism experiments~\cite{Emori:apl24}, although such detection requires the higher time resolution than that of electron-phonon scattering processes~\cite{Giustino:rmp17}.

\begin{figure}[t]
\centering
\includegraphics[width=\columnwidth]{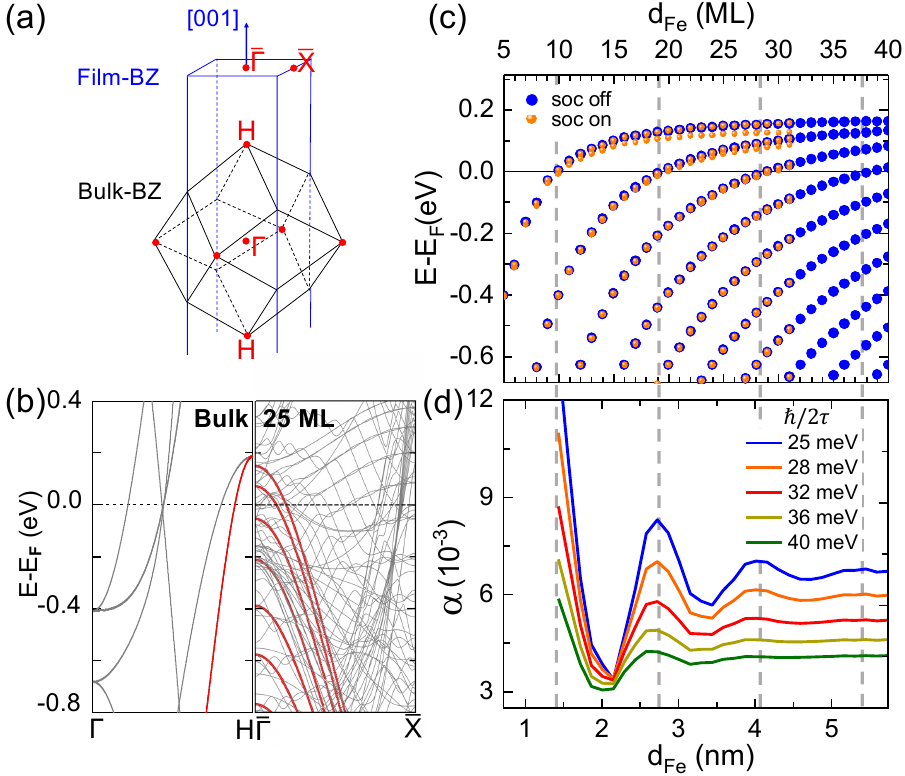}
\caption{(a) Bulk BZ of bcc lattice (black) and 2D BZ of (001) films (blue). (b) Comparison of energy bands of bulk Fe and a 25-ML-thick Fe film. The bands of $\Delta_5^{\uparrow}$ character are marked by red color. SOC is neglected for clarity. (c) Calculated energies of the discrete $\Delta_{5}^{\uparrow}$ QWSs at the $\bar{\Gamma}$ point as a function of Fe thickness. Including SOC results in negligible changes in the energies of QWSs. (d) Calculated damping as a function of Fe thickness with various scattering rates, which are larger than those for bulk calculation to capture the surface scattering effect. The grey dashed lines in (c) and (d) identify the thicknesses, at which the $\Delta_{5}^{\uparrow}$ QWSs are located at $E_F$ corresponding to the maximum values of oscillatory $\alpha$.}
\label{fig3}
\end{figure}

%%%%%%%%10%%%%%%%%20%%%%%%%%30%%%%%%%%40%%%%%%%%50%%%%%%%%60%%%%%%%%70%%%%%%%%80
{\it\color{red}Distinguishing band contributions via QWSs.---}All the partially occupied $d$ bands at the Fermi level contribute to the Gilbert damping and distinguishing them requires finding the fingerprint of each band. QWSs arising from spatial confinement in thin films usually induces quantum oscillation in the DOS at $E_F$ and many related physical properties. The oscillation period depends on the Fermi wave vector $k_F$ of the specific band~\cite{Smith:PRB94}. The quantum oscillation of tunneling magnetoresistance in ultrathin Fe films shows the period of approximately 2 monolayers (MLs) in agreement with $k_F$ of $\Delta_1^\uparrow$ band because this band dominates the tunneling conductance~\cite{Lu:PRL05,Nozaki:PRL06,Niizeki:PRL08,Tao:PRL15}. The magnetocrystalline anisotropy of Fe films is mainly influenced by the $\Delta_2^{\uparrow(\downarrow)}$ bands and hence its quantum oscillation has the period of 5 to 6 MLs~\cite{Wu:PRL05,Li:PRL09}. In the same manner, the QWSs in Fe films shall induce the quantum oscillation of Gilbert damping, allowing us to verify the dominant band. Fig.~\ref{fig3}(a) shows the correspondence of the bulk and film BZs and the calculated energy bands of bulk Fe and a thin film with the thickness of $d_{\rm Fe}=$25 MLs are plotted in Fig.~\ref{fig3}(b). The continuous $\Delta_5^{\uparrow}$ band (red line) along $\Gamma\rm H$ in bulk Fe are discretized into the subbands (red dots) in the thin film, which are located around $\bar{\Gamma}$ point in the 2D BZ. The tops of $\Delta_5^{\uparrow}$ subbands at $\bar{\Gamma}$, which have the largest spectral function due to the zero dispersion, are plotted in Fig.~\ref{fig3}(c) as a function of $d_{\rm Fe}$. With increasing $d_{\rm Fe}$, these QWSs are gradually lifted upwards and cross $E_F$ periodically.

We calculate the intraband contribution using Eq.~\eqref{eq:tcm} with two approximations~\cite{SM}: (1) only the spectral functions at the $\bar{\Gamma}$ point is taken into account; and (2) the $k$-dependent intraband matrix element $\vert\Gamma^{-}_{nn}(\mathbf k)\vert^2$ is replaced by the average over the Fermi sheet $\langle\vert\Gamma^-_n\vert^2\rangle$. Then, Eq.~\eqref{eq:tcm} is simplified as 
\begin{equation}
\alpha=\frac{\pi\hbar\gamma}{\mu_0M_s}\sum_{n}\langle\vert\Gamma^-_n\vert^2\rangle\int dE\,A_{n\bar{\Gamma}}^2(E)\eta(E).\label{eq:2d}
\end{equation}
The calculated Gilbert damping is plotted in Fig.~\ref{fig3}(d) as a function of $d_{\rm Fe}$, where the scattering rates are larger than those in Fig.~\ref{fig1}(c) to account for the surface scattering effect. At low scattering rate $\hbar/2\tau=25$~meV to 32~meV, $\alpha$ exhibits strong oscillations and the maximum values of $\alpha$ appears at the thickness with the QWS across $E_F$, as indicated by the grey dashed lines.  The oscillation period~\cite{Smith:PRB94,Wu:PRL05} is determined by $\pi/(k_{\rm BZ}-k_F)\approx9$~MLs, where $k_F$ is the Fermi wave vector of the $\Delta_5^\uparrow$ bands and $k_{\rm BZ}$ represent the distance from $\Gamma$ to H in the bulk BZ. The oscillation period is another evidence for the dominant contribution of the $\Delta_5^\uparrow$ bands to $\alpha$. As the scattering rate increases, the oscillation strength gradually decreases and almost vanishes at $\hbar/2\tau=40$~meV. 

%%%%%%%%10%%%%%%%%20%%%%%%%%30%%%%%%%%40%%%%%%%%50%%%%%%%%60%%%%%%%%70%%%%%%%%80
\begin{figure}[t]
\centering
\includegraphics[width=0.9\columnwidth]{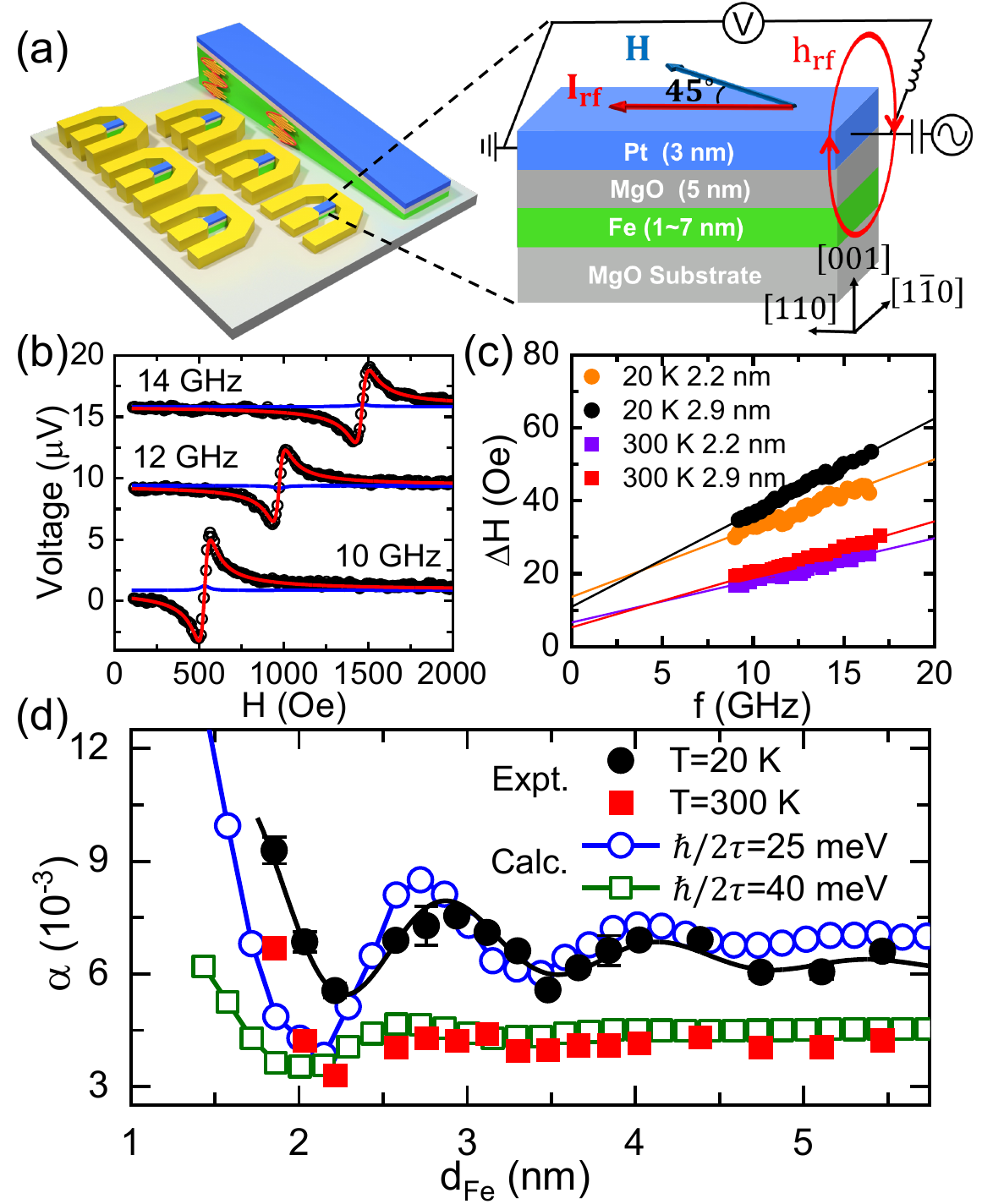}
\caption{(a) Schematic illustration of the experimental sample. Microwave current ($I_\mathrm{rf}$) is injected into the sample along Fe[110] via the Au coplanar waveguides (yellow shapes) positioned on top of the Fe samples (blue stripes). This setup generates a microwave field in the Fe layer perpendicular to the strips. (b) Representative field-dependent spin rectification voltage measured on a 2.2-nm-thick Fe film at 20 K with frequencies of $f$=10, 12, and 14~GHz. The red and blue lines represent fitting curves for the antisymmetric and symmetric components, respectively. (c) Linewidth ($\Delta H$) as a function of frequency for both 2.2-nm-thick and 2.9-nm-thick samples. The solid lines depict linear fitting. (d) Measured Gilbert damping as a function of Fe thickness (solid symbols) compared with the theoretical calculations (empty symbols). The black solid line illustrates the quantum oscillation of the experimental values at low temperature. 
}
\label{fig4}
\end{figure}

{\it\color{red}Experimental confirmation of damping oscillation.---}The single-crystal Fe films were grown on MgO(001) substrates using molecular beam epitaxy in an ultrahigh vacuum chamber with a base pressure of $2\times10^{-10}$ Torr. The epitaxial growth of wedge-shaped Fe films was carried out at room temperature, as sketched in Fig.~\ref{fig4}(a), with crystalline quality characterized by reflection high-energy electron diffraction and X-ray diffraction~\cite{SM}. Subsequently, 5-nm-thick MgO and 3-nm-thick Pt films were deposited on top of the Fe wedge.  

The samples were patterned into a series of strips with the size of $200\,\mu\mathrm m\times15\,\mu\mathrm m$ using standard ultraviolet lithography and argon ion etching. Subsequently, coplanar waveguides containing the Au(150 nm)/Cr(30 nm) bilayer were fabricated through magnetron sputtering. The microwave current in the Pt layer generates a radio frequency field, exciting FMR of the underlying Fe layer. During the measurement, an in-plane magnetic field was applied along Fe[100], 45$^\circ$ away from the stripes. The spin rectification due to the microwave current and the oscillating anisotropic magnetoresistance of Fe results in an antisymmetric Lorentzian line shape for the measured voltage~\cite{Hu:PRL07,Hu:PhysRep16}, as shown in Fig.~\ref{fig4}(b). The resonant magnetic field $H_\mathrm{r}$ and the linewidth $\Delta H$ were determined by fitting the voltage using the Lorentzian function $V(H)=V_a\frac{\Delta H\left(H-H_{\rm r}\right)}{\left(H-H_{\rm r}\right)^2+\Delta H^2}+V_s\frac{\Delta H^2}{\left(H-H_{\rm r}\right)^2+\Delta H^2}$. The antisymmetric and symmetric components are plotted in Fig.~\ref{fig4}(b) by the red and blue lines, respectively. The symmetric contribution is negligible indicating that MgO is enough thick to block spin pumping~\cite{Pap:APL17}. Fig.~\ref{fig4}(c) presents the typical linear dependence of the linewidth $\Delta H$ on the FMR frequency and the Gilbert damping $\alpha$ was extracted through linear fitting, $\Delta H=\Delta H_0+4\pi \alpha f/\gamma$. Here the gyromagnetic ratio $\gamma=g\mu_0\mu_B/\hbar$ and the $g$-factor are determined by fitting the $f-H_\mathrm{r}$ curves to the Kittel equation~\cite{Kittel:PhysRev48}. While the fitted $g$-factor is nearly thickness-independent~\cite{SM}, the variation of the slopes with thickness in Fig.~\ref{fig4}(c) at 20 K arises from different $\alpha$. Additional analysis confirms that the change in $\alpha$ is not related to the inhomogeneous linewidth $\Delta H_0$ or two-magnon scattering~\cite{SM,TMS1,TMS2}.

The measured Gilbert damping is plotted in Fig.~\ref{fig4}(d) as a function of film thickness. At low temperature, $\alpha$ exhibits a significant oscillation (black solid circles). Both the oscillation period and the thicknesses for the maximum damping are in quantitative agreement with the calculations using the low scattering rate $\hbar/2\tau=25$~meV (blue empty circles). At 300 K, the measured $\alpha$ (red solid squares) is smaller and the oscillation becomes very weak, in agreement with the calculated values using a large scattering rate. The measurement is performed for another independent sample resulting in the same conclusions~\cite{SM}. All the experimental results are in very good agreement with the calculations confirming the dominant contribution of the $\Delta_5^\uparrow$ bands to the Gilbert damping of Fe. %In these bands with pure spin character at equilibrium, the spin dissipation arises mainly from the orbital excitations.
	
%%%%%%%%10%%%%%%%%20%%%%%%%%30%%%%%%%%40%%%%%%%%50%%%%%%%%60%%%%%%%%70%%%%%%%%80
{\it\color{red}Conclusion.---}Using first-principles calculations, we have discovered that the Gilbert damping of Fe is dominated by the orbital excitations of the $\Delta_5^\uparrow$ bands. The variation of orbital angular momentum modifies the spin component through SOC. The dominant role of the  $\Delta_5^\uparrow$ bands is unambiguously confirmed by the excellent agreement between low-temperature experiment and calculations of quantum oscillation of Gilbert damping in ultrathin Fe(001) films. Our findings not only reveals the nature of the recently reported ultralow damping in Fe-rich alloys, but also provides an efficient and general way for controlling Gilbert damping by selectively shifting the relevant bands away from the Fermi level via doping. In two-dimensional magnetic materials, bands shifts can be achieved by gating~\cite{Burch:Nat18,Gong:Sci19,Yang:prb22}. The discovery of orbital excitations in the magnetization dissipation would be helpful for studying transport and dissipation of orbital currents~\cite{Ding:prl22,Choi:nat23,Seifert:natnano23,Lyalin:prl23,Sala:prl23,Sohn:prl24,Rang:prb24}. 
	
%%%%%%%%10%%%%%%%%20%%%%%%%%30%%%%%%%%40%%%%%%%%50%%%%%%%%60%%%%%%%%70%%%%%%%%80
\acknowledgements
We thank valuable suggestions and comments provided by Professor H. F. Ding and Professor B. Heinrich. The work was supported by the National Key Research and Development Program of China (2022YFA1403300 and 2024YFA1408500), the National Natural Science Foundation of China (Grants No. 12174028, No. 12374101, No. 11974079, and No. 12274083), the Shanghai Municipal Science and Technology Major Project (Grant No. 2019SHZDZX01), the Shanghai Municipal Science and Technology Basic Research Project (No. 22JC1400200 and No. 23dz2260100), and the Open Fund of the State Key Laboratory of Spintronics Devices and Technologies (Grants No. SPL-2408).
	
%%%%%%%%10%%%%%%%%20%%%%%%%%30%%%%%%%%40%%%%%%%%50%%%%%%%%60%%%%%%%%70%%%%%%%%80

\end{document}